\documentclass[10pt,conference,onecolumn]{IEEEtran}
\IEEEoverridecommandlockouts
\usepackage{comment}
\usepackage[T1]{fontenc}
\usepackage{algorithm,amsmath}
\usepackage{algpseudocode}
\usepackage[cmintegrals]{newtxmath}
\usepackage{cite}
\usepackage[utf8]{inputenc}
\usepackage[english]{babel}
\usepackage[]{verbatim}
\usepackage{syntonly}
\usepackage{textcomp}
\usepackage{amsmath}
\usepackage{ upgreek }
\usepackage{subcaption}
\usepackage{graphicx} 
\usepackage{varioref}
\usepackage[noabbrev]{cleveref}
\usepackage{textcomp,xcolor}
\usepackage[figurename=Fig.]{caption}
\usepackage{float}

\IEEEoverridecommandlockouts\IEEEpubid{\makebox[\columnwidth]{ 978-1-6654-5975-
4/22~\copyright~2022 IEEE \hfill} \hspace{\columnsep}\makebox[\columnwidth]{ }}
\begin{document}

	\title{Beyond-Cell Communications via HAPS-RIS\\
		\thanks{This  research  has  been  supported in part by a scholarship from King AbdulAziz
			University, Saudi Arabia, and in part by
			Huawei Canada Company Ltd.}}
			
	
		    \author{\IEEEauthorblockN{Safwan Alfattani\IEEEauthorrefmark{1}\IEEEauthorrefmark{2},
			Animesh Yadav\IEEEauthorrefmark{4}, Halim Yanikomeroglu,\IEEEauthorrefmark{3} and Abbas Yonga{\c{c}}oglu\IEEEauthorrefmark{2}}
		\IEEEauthorblockA{\IEEEauthorrefmark{1}King AbdulAziz
			University, Saudi Arabia, \IEEEauthorrefmark{2}University of Ottawa,  Canada, \IEEEauthorrefmark{3}Carleton University,  Canada, and\\ \IEEEauthorrefmark{4}Minnesota State University, Mankato, MN, USA\\
			Email: smalfattani@kau.edu.sa,
			animesh.yadav@mnsu.edu,
			halim@sce.carleton.ca, yongac@uottawa.ca} }	
	\maketitle
	
	\begin{abstract}
		The ever-increasing number of users and new services in urban regions 
		can lead terrestrial base stations (BSs) to become overloaded and, consequently, some users to go unserved.
		 Compounding this, users in urban areas can face severe shadowing and blockages, which means that some users do not receive a desired quality-of-service (QoS).
 Motivated by the energy and cost benefits of reconfigurable intelligent surfaces (RIS) and the advantages of high altitude platform station (HAPS) systems, including their wide footprint and strong line-of-sight (LoS) links, 
we propose a solution to support the stranded users using the RIS-aided HAPS. 
Particularly,
 we propose to support the stranded users by a dedicated control station (CS) via a HAPS equipped with RIS (HAPS-RIS). Through this approach,  users are not restricted from being supported by the cell they belong to; hence, we refer to this approach as \textit{beyond-cell} communication. 
 As we demonstrate in this paper, 
   \textit{beyond-cell} communication works in tandem with  legacy terrestrial networks to support uncovered or unserved users. Optimal transmit power and RIS unit assignment strategies for the users based on different network objectives are introduced. Numerical results demonstrate the benefits of the proposed \textit{beyond-cell} communication approach. Moreover, the results provide  insights into the different optimization objectives and their interplay with minimum QoS and network resources, such as transmit power and the number of RIS reflecting units.
	\end{abstract}

	\section{Introduction}
	One of the main goals of 
	\textcolor{black}{future} wireless network is to provide ubiquitous connectivity  (i.e., wireless connectivity to everyone and everything, everywhere, every time at an affordable rate) \cite{6G_Frontiers}. Ubiquitous connectivity can be achieved in urban areas through the dense deployment of base stations (BSs), including small cells. However, with  increasing numbers of users 
	and hampered by severe shadowing, blockages, and non-line-of-sight (NLoS) links, even ultra dense networks
	 cannot support all users in an urban area. Further, deploying a large number of BSs inevitably leads to high capital expenditures (CAPEX) and operational expenditures (OPEX) 
	\cite{RANs_survey}.

	Recent studies have proposed deploying reconfigurable intelligent surfaces (RIS) around BSs
	 \textcolor{black}{as an energy-efficient solution to overcome  severe shadowing and blockage effects \cite{kishk2020exploiting}. 
	 An RIS is a reflecting surface built from a massive number of tiny reflecting units \cite{alfattani2021aerial,di2020smart,liu2021reconfigurable}. Each reflecting unit is 
	 controllable,
	 and thus 
	 it can 
	 focus the impinging  signals 
	 in a desired direction in a nearly passive way. 
	  Thus, RIS are  energy-efficient alternative to active antenna architecture such as relays \cite{alfattani2021aerial,di2020smart}.} 
	 \textcolor{black}{However, the deployment of RIS in terrestrial networks involves several challenges, including inflexible deployment and weak wireless channel conditions due to shadowing, blockages and NLoS links.} 
	 In addition, \textcolor{black}{dynamic and unpredictable} network traffic 
	 generates \textcolor{black}{unprecedented} data rate demands, which can overload some  BSs. Accordingly, even an optimized deployment of BSs and RIS in urban areas might  be unable to cope with the dynamic demands. 
	
	To overcome these issues, 
	we recently proposed integrating RIS with  non-terrestrial networks (NTN) in  previous works \cite{alfattani2021aerial,kurt2021vision}.
	\textcolor{black}{ Due to the limited energy on aerial platforms, integrating RIS with NTN is more appealing than integrating them in terrestrial networks.}
	In these works, \textcolor{black}{we also discussed several benefits of RIS-aided NTN, including energy and cost savings, favorable wireless channel conditions,} strong LoS links, a wider coverage area, and flexible placement. 
	In another work \cite{alfattani2021link}, we also provided a detailed link budget analysis
	 of 
	 different RIS-aided aerial platforms, such as
	  unmanned aerial vehicles (UAVs), high altitude platform station (HAPS) nodes, and low Earth orbit (LEO) satellites, and we compared that to RIS-aided terrestrial networks.
	
	As we also showed in \cite{alfattani2021link}, the typical large size of a HAPS allows the accommodation of a large number of \textcolor{black}{reflecting units}. Thus, it 
can outperform other aerial RIS-aided systems, such as UAV-RIS.
	 Therefore, \textcolor{black}{ motivated by the aforementioned findings, 
	  we propose in this paper} a novel \textit{beyond-cell} communications approach involving an \textcolor{black}{HAPS equipped with RIS}  (HAPS-RIS). This approach works in tandem with legacy terrestrial networks \textcolor{black}{by offering service to} unsupported users whose quality-of-service (QoS) requirements cannot be fulfilled by \textcolor{black}{legacy} networks. 
	  In our proposed scheme, unsupported users are connected to a dedicated control station (CS) via a HAPS-RIS.
	   The main contributions in this paper include, but are not limited to, the following:
	\begin{itemize}
		\item  We formulate three novel optimization problems, including throughput maximization, worst user rate maximization, and \textcolor{black}{reflecting units} usage minimization to design optimal power and RIS unit assignment strategies for the users supported through \textit{beyond-cell} communications.
		\item We present thorough numerical simulation results, which indicate that  \textit{beyond-cell} communications 
		complement legacy terrestrial networks well and
		  enhance the total number of users served. The percentage of users that satisfy the QoS requirements increases while the required number of terrestrial BSs decreases. 
		\item We also provide  important insights about the different allocation schemes. By conceding a small degradation (1\%) in the system sum rate,  the 
		worst user rate maximization-based allocation scheme maximizes the fairness among the users and increases the rate of the worst user ($\sim$15\%). Moreover, by increasing the CS
		power by 1 dB, the size of the HAPS or the required number of reflectors is decreased by $\sim$11\%.
	\end{itemize}
	
	The remainder of the paper is organized as follows.  Section \ref{Sec:model} presents the system model for all 
	 user equipment (UE), whether connected
	 to a BS or supported by a HAPS-RIS.
	 Section \ref{Sec:Problem_form} details the problem formulations for different system objectives. The proposed solutions are discussed in Section \ref{Sec:solutions}, while the numerical results are presented and discussed in Section \ref{Sec:Results}. Finally, Section \ref{Sec:conclstion}  concludes the paper.
	
	\begin{figure}[h]
		\centering
		\includegraphics[scale=0.9]{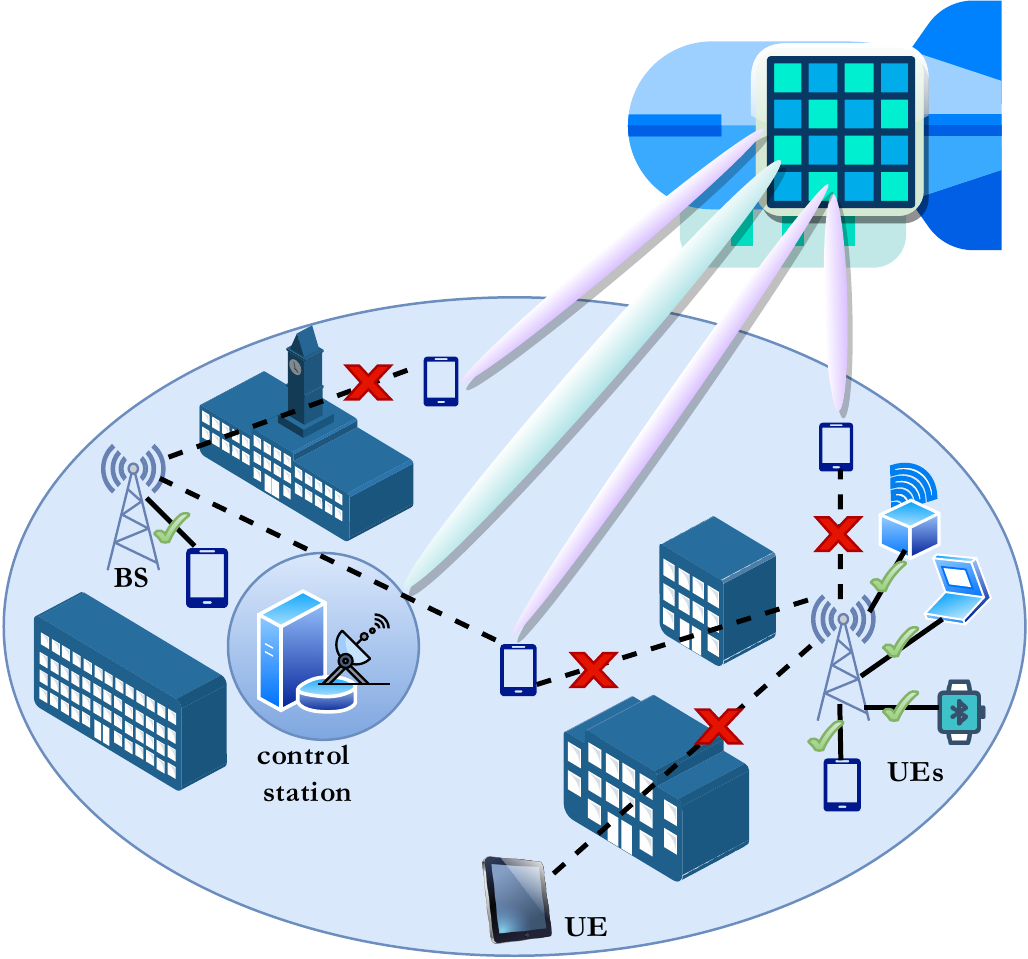}
		\caption{System model for HAPS-RIS  beyond-cell communications.}
		\label{fig:model}
	\end{figure}
	
	\section{System Model} \label{Sec:model}
	We consider a downlink transmission scenario in an integrated terrestrial and non-terrestrial wireless network consisting of a single HAPS with a coverage area that includes $L$  base stations (BSs) and $K$ single-antenna UEs (e.g., smart phones, sensing nodes or Internet-of-Things devices, etc.). The HAPS is located in the center of the coverage area at an altitude of 20 km\footnote{\textcolor{black}{Generally, HAPS systems are quasi-stationary \cite{kishk2020exploiting}, and hence, their mobility is ignored in this model.}}, and it is equipped with the RIS with a total of $N_{\rm max}$  reflecting units. 
	\textcolor{black}{Since the antennas of  BSs are down-tilted to serve terrestrial UEs, we consider  the HAPS coverage area also includes a dedicated high-directional antenna gain transceiver, known as the ground CS.  The CS is connected to the core network with the primary aim of supporting unserved UEs via the HAPS-RIS. Further, the CS also manages the association between the BSs and the UEs and configures the RIS.}
	
	The environment for this scenario is an urban one
	where $K$ UEs are uniformly distributed in the HAPS coverage area. In contrast, the BSs are optimally placed to maximize the number of connected UEs. We adopt 
	\textcolor{black}{Lloyd's} algorithm for optimizing the 
	placement of BSs.
	 The algorithm aims to minimize the Euclidean distances between the UEs and the BSs; 
	  this distance is the factor that most affects the quality of the channel links\footnote{\textcolor{black}{This algorithm is also known as a $k$-means clustering algorithm \cite{lu2016statistical}.}}. In an urban environment, wireless signals suffer severe blockages and shadowing \cite{3gpp38901study}. Accordingly, some UEs might not be associated with any BSs due to poor channel quality. Moreover, some 
	  UEs might not have service
	   when BSs are fully loaded. \textcolor{black}{Thus, the UEs in the system can be divided into two groups. The first group includes the UEs that 
	can be served by  BSs; the second group 
consists of the remaining UEs that
	can be served by the CS via HAPS-RIS.}
	
	\subsection{Within-Cell Connection: UEs to Terrestrial BSs}
	In this work, we employ the orthogonal frequency division multiplexing (OFDM) transmission scheme, where the entire terrestrial system bandwidth $B_{\rm BS}$ Hz is divided into equal subcarriers each of bandwidth $B_{\rm UE}$ Hz. \textcolor{black}{Without loss of generality,} the UEs in each cell are allowed to use only one subcarrier to transmit their data. However, the UEs from other cells can use the same subcarrier. Thus, there will be inter-cell interference but no intra-cell interference.
	
	Accordingly, the received signal at  UE $k$ from  BS $l$ on a given subcarrier can be written as
	\begin{equation}
	\label{eq:Rx_direct_signal}
	y_{kl}=\sqrt{P^{\rm BS}_{kl}} h_{kl}x_{kl}+ \sum_{l^{\prime}=1,l^{\prime}\neq l}^{L} \sqrt{P^{\rm BS}_{kl^{\prime}}} h_{kl^{\prime}} x_{kl^{\prime}} +w_{kl},
	\end{equation}
	where $P^{\rm BS}_{kl}$ and $x_{kl}$ denote the transmitted power and the transmitted symbol, respectively.  $w_{kl}$ denotes  the additive white Gaussian noise (AWGN) with zero mean and power spectral density $N_0$. $h_{kl}$ denotes the channel coefficient between  UE $k$ and BS $l$ on a given subcarrier, and it can be expressed as
	\begin{equation}
	h_{kl} = \sqrt{G^{\rm BS}_{l}G_{k}^{\rm UE} (\textsf{PL}_{kl}^{\rm BS})^{-1}}, 
	\end{equation}
	where $G^{\rm BS}_{l}$ and $G_k^{\rm UE}$ denote the antenna gain of  BS $l$ and UE $k$, respectively. $\textsf{PL}_{kl}^{\rm BS}$ denotes the path loss of the channel between  BS $l$ and  UE $k$.  
	Now, the achievable signal-to-interference-plus-noise ratio (SINR) at  UE $k$ served by  BS $l$ can be written as
	\begin{equation}\label{eq:SINR}
	\gamma_{kl}= \frac{P_{kl}^{\rm BS} \left| h_{kl}  \right|^2} {\sum_{l^{\prime}=1,l^{\prime}\neq l}^{L} P_{kl^{\prime}}^{\rm BS}\left| h_{kl^{\prime}}\right|^2  + N_0 B_{\rm UE}}.
	\end{equation}
	The corresponding achievable rate in bits per seconds (bps) between  BS $l$ and  UE $k$ can be written as
	\begin{equation}\label{eq:R_K1}
	R_{kl} = B_{\rm UE} \log_2(1+\gamma_{kl}).
	\end{equation}
	
	For a UE to be directly associated with a BS $l$, the data rate between them should be above the  minimum required data rate, i.e., $R_{kl} \geq R_{\rm min}$. Let $\mathcal{S}_k \subset  \{R_{k1},...,R_{kL}\}$ denote the set of data rates between  UE $k$ and $L$ BSs that have a data rate higher than the minimum required rate $R_{\rm min}$.
	The UE $k$ is associated with the BS $l$ with the highest data rate in the set $\mathcal{S}_k$, i.e., ($l=\displaystyle\max_{\boldsymbol{l}}  \mathcal{S}_k$). Also, let $\mathcal{M}_{l} \subset  \{1,...,k\}$ denote the set of UEs with the \textcolor{black}{best channel} toward BS $l$, and the cardinality of  the set is denoted as $\left|\mathcal{M}_l\right|$.
	
	For the UE $j, j\neq k$ with  $\mathcal{S}_{j}=\{\phi\}$ (i.e., $ R_{jl} < R_{\rm min}$, $\forall~ l \in \{1,...,L\}$), its communication is supported by the CS through HAPS-RIS. Also, when $\left|\mathcal{M}_l\right| > (B_{\rm BS}/B_{\rm UE}) $, we declare the BS $l$ as fully loaded or its capacity as fully utilized. Hence, we drop UEs with the lowest channel gain to be served by the CS via HAPS-RIS until $\left|\mathcal{M}_l\right| \leq (B_{\rm BS}/B_{\rm UE}) $. 
	Accordingly, we denote the set of $K_1$ UEs supported by direct links from the BSs (\textit{within-cell} communications) with $\mathcal{K}_1 = \{1,\ldots,K_1\}$. Similarly, we denote the set of $K_2$ UEs supported by the CS through HAPS-RIS (\textit{beyond-cell} communications) with $\mathcal{K}_2 = \{1, \ldots, K_2\}$. 
	
	\subsection{Beyond-Cell Connection: UEs to CS via HAPS-RIS}
	The unsupported UEs that cannot form a direct connection with the terrestrial BS will be served by the CS via HAPS-RIS. We assume that the CS serves the UEs in  set $\mathcal{K}_2$ using the OFDM protocol, and hence, there will be no inter-UE interference. Further, both \textit{within-cell} and \textit{beyond-cell} communications occur in two orthogonal frequency bands, while keeping $B_{\rm UE}$ same for both types of connection. Consequently, there will be no interference between  \textit{within-cell} UEs belonging to set $\mathcal{K}_1$ and \textit{beyond-cell} UEs belonging to set $\mathcal{K}_2$. 
	
	Accordingly, the received signal at UE $k \in \mathcal{K}_2$  can be expressed as
	\begin{equation}
	\label{eq:Rx_signal}
	y_{k}=\sqrt{P^{\rm CS}_{k}} h_{k}  \mathrm{\Phi}_{k} \;x_{{k }}+w_{k},
	\end{equation}
	where $P^{\rm CS}_{k}$ and $w_{k}$ denote the transmit power and the zero-mean AWGN of UE $k$, respectively. $h_{k}$ denotes the \textit{effective} channel gain from the CS to the HAPS-RIS and from the HAPS-RIS to UE $k$, and it is given by \cite{alfattani2021link}{\footnote{\textcolor{black}{Here, we consider negligible difference in channels between UE $k$ and RIS reflecting units, since they are dominated by LoS links.}}} 
	
	\begin{equation}
	h_{k} = \sqrt{G^{\rm CS}_{\rm t} G_{\rm r}^{k} (\textsf{PL}
^{\rm{CS-HAPS}-\textit{k}})^{-1}},    
	\end{equation}
	where $G^{\rm CS}_{\rm t}$ denotes the 
	transmit antenna gain of  the control station, and $G_{\rm r}^{k}$ is the receiver antenna gain of UE  $k$. 
	$\textsf{PL}
^{\rm{CS-HAPS}-\textit{k}}$ represents the cascaded path loss between the control station and the HAPS (i.e., $\textsf{PL}
^{\rm CS- HAPS}$), and between the HAPS and UE $k$ (i.e., $\textsf{PL}
^{\text{HAPS-}k}$).
	Furtehr, $\mathrm{\Phi_{k}}$ represents the reflection gain of the RIS corresponding to UE $k$, and is expressed as
	\begin{equation}
	\mathrm{\Phi}_{k}=\sum_{i=1}^{N_{k}} \rho_i e^{-j  \left(\phi_i - \theta_{i}-\theta_{k}\right)},
	\end{equation}
	where $\rho_i$ denotes the reflection loss corresponding to reflector unit $i$, while $\theta_{i}$ and $\theta_{k}$ represent the corresponding phases between reflector unit $i$ and both the control station and UE $k$, respectively. $\phi_i$ represents the adjusted phase shift of reflector unit $i$, while $N_{k}$ represents the total number of reflecting units allocated to UE $k$. The RIS reflecting units are divided among the UEs because each UE uses different subcarrier.
	
	Accordingly, the signal-to-noise ratio (SNR) at UE $k$ can be written as
	\begin{equation}\label{eq:SNR}
	\gamma_{k}= \frac{P_{k}^{\rm CS} \left| h_{k}  \mathrm{\Phi}_{k}\right|^2} { N_0 B_{\rm UE}},
	\end{equation}
	and the achievable rate of UE $k \in \mathcal{K}_2$ can be expressed as
	\begin{equation}\label{eq:R_K2}
	R_{k} = B_{\rm UE} \log_2(1+\gamma_{k}).
	\end{equation}
	\section{Problem Formulation}\label{Sec:Problem_form}
	In this section, we discuss three resources (available power at the CS and reflecting units at the HAPS) allocation schemes for the UEs assisted by the \textit{beyond-cell} communication, and accordingly, formulate three optimization problems.
	
	\subsection{Sum Rate (Throughput) Maximization }
	The goal of this problem is to support all the $K_2$ UEs and maximize their sum rate by optimally allocating the reflecting units and transmitting power to all the UEs. Thus, the formulation becomes
{\color{black}
\begin{IEEEeqnarray*}{lcl}\label{eq:sum_rate_max}
&\underset{\mathrm{\Phi}_{k},N_{k},P_{k}^{\rm CS}}{\max}\,\, &  \sum_{k=1}^{K_2}R_{k}  \IEEEyesnumber \IEEEyessubnumber* \label{eq:maxR}\\
&\text{s.t.} & L \leq L_{\rm max}, \label{p1_throu:c1}\\
&& R_{k} \geq R_{\rm th}, \enspace \forall k=1,2, \ldots, K_2, \label{p1_throu:c2}\\
&& \sum_{k=1}^{K_2}  N_{k} \leq N_{\rm max}, \label{p1_throu:c3}\\
&&  \theta_{n} \in \{0, 0+b, \ldots, 2\pi\}, \: \text{ \footnotesize $\forall n=1, \ldots, N_{ \rm max}$,} \label{p1_throu:c4}\\ 
&&\sum_{k=1}^{K_2} P_{k}^{\rm CS} \leq P_{\rm max}^{\rm CS}, \label{p1_throu:c5}\\
&&N_{k} \in \{N_{k, \rm min},N_{k, \rm min}+1,\ldots, N_{k, \rm max}\},\label{p1_throu:c6}\\ &&P_{k, \rm min}^{\rm CS} \leq P_{k}^{\rm CS} \leq P_{k, \rm max}^{\rm CS},\label{p1_throu:c7}
\end{IEEEeqnarray*}}
where (\ref{p1_throu:c1}) limits  the number of terrestrial BSs in the area; and thus, limits the expenditure incurred by network operators. \textcolor{black}{Note that, the value of $K_2$ is dependent on the value of $L$. Higher the value of $L$, lower is the value of $K_2$.} Constraint (\ref{p1_throu:c2}) ensures that the minimum required rate is achieved by each UE. Constraint (\ref{p1_throu:c3}) guarantees that the total number of allocated reflecting units to all UEs is less than the maximum number available at the HAPS. In practice, the value of $N_{\rm max}$ is dependent on the HAPS size. Constraint (\ref{p1_throu:c4}) \textcolor{black}{specifies the discrete range of the adjustable phase shifts for the reflectors, where $b$ is dependent on the resolution of the phase shift.}  
Constraint (\ref{p1_throu:c5}) ensures the total allocated power for each UE does not exceed the maximum power\textcolor{black}{, $P_{\rm max}^{\rm CS}$,} of the control station. \textcolor{black}{Constraints (\ref{p1_throu:c6})-(\ref{p1_throu:c7}) ensure feasible and fair allocation of both reflecting units and CS power, respectively, where $N_{k, \rm min}, N_{k, \rm max}$, $P_{k, \rm min}^{\rm CS}$, and $P_{k, \rm min}^{\rm CS}$ denote the minimum and maximum allocated  RIS units and power per UE.}
	\subsection{Minimum Rate Maximization (Max-Min Rate)}
Sum rate (throughput) maximization based allocation scheme might be biased toward UEs with better channel links. Therefore, for fair resource allocation, we consider in this subsection the problem of maximizing the minimum rate among all UEs. Accordingly, the max-min fairness problem can be formulated as
\begin{IEEEeqnarray*}{lcl}\label{eq:max_min_R_general}
&\underset{\mathrm{\Phi}_{k},N_{k},P_{k}^{\rm CS}}{\text{maximize}}\,\, & \min_{k=1,\dots, K_2}\, R_{k}  \IEEEyesnumber \IEEEyessubnumber*\label{eq:max_min_R}\\
&\text{s.t.} & \eqref{p1_throu:c1}-\eqref{p1_throu:c7}.\label{p2:c1}
\end{IEEEeqnarray*}
	\subsection{Reflecting Units Minimization}
	Despite being passive in nature, the RIS reflecting units consume energy for control and configuration \cite{alfattani2021aerial,di2020smart}. The energy consumption might be significant for HAPS equipped with a large number of reflectors. For a cost-effective deployment of HAPS in terms of on-board energy consumption and size of HAPS, it is essential to minimize the total number of RIS reflecting units. Accordingly, the RIS reflecting units minimization problem can be formulated as \begin{IEEEeqnarray*}{lcl}\label{eq:min_N_general}
    &\underset{\mathrm{\Phi}_{k},N_{k},P_{k}^{\rm CS}}{\min} \, \, & \sum_{k=1}^{K_2}  N_{k}  \IEEEyesnumber \IEEEyessubnumber*\label{eq:min_N}\\
    &\text{s.t.} & \eqref{p1_throu:c1}-\eqref{p1_throu:c7}.\label{p2:c1}
    \end{IEEEeqnarray*}

	\section{Proposed Solution} \label{Sec:solutions}
	In this section, we discuss the solution approaches for the aforementioned optimization problems.
	\subsection{Sum Rate Maximization}
	In problem \eqref{eq:sum_rate_max},  constraint (\ref{p1_throu:c1}) is determined by an expenditure analysis of the network deployment, and typically it is selected as $L=L_{\rm max}$ to maximize the percentage of UEs with direct connections to BSs.
	It should be noted that the variables $N_{k}$ and $\theta_n$ are discrete variables; and thus (\ref{eq:sum_rate_max}) becomes 
	computationally challenging to solve as it requires employing heuristic discrete optimization algorithms. 
	However, for a large RIS area, $N_{k}$ can be approximated as a continuous variable. This approximation is substantiated by the fact that each UE is  allocated a subarea of the RIS, and the 
	total area allocated to all UEs
	 is approximately equivalent to the total RIS area.
	Similarly, $\theta_n$ in (\ref{p1_throu:c4}) practically has a range of discrete phase shifts. However, 
	several works have shown that close
	 to optimal continuous performance can be achieved even with low resolution discrete phase shifts \cite{rivera2022optimization}. Therefore, $\theta_n$ can be approximated as a continuous variable.  
	Moreover, \textcolor{black}{for large values of $\gamma_{k}$}  the rate of  UE $k$, given in (\ref{eq:R_K2}), can be approximated as $R_{k} \approx \textcolor{black}{B_{\rm UE}} \log_2(\gamma_{k}).$
	Accordingly, the sum rate of the set $\mathcal{K}_2$ UEs is given as 
	\begin{equation}
	\sum_{k=1}^{K_2}R_{k} \approx  \textcolor{black}{B_{\rm UE}} \log_2\Bigg(\prod_{k=1}^{K_2}\gamma_{k}\Bigg).
	\end{equation}
	Moreover, the rate constraint (\ref{p1_throu:c2}) of each UE can be written in terms of its SNR.
	Accordingly, problem (\ref{eq:sum_rate_max}) can be  reformulated as
	\begin{IEEEeqnarray*}{lcl}\label{eq:maxR_2}
    &\underset{\mathrm{\Phi}_{k},N_{k},P_{k}^{\rm CS}}{\min} \, \, & \dfrac{1}{\prod_{k=1}^{K_2} \gamma_{k}}  \IEEEyesnumber \IEEEyessubnumber*\label{eq:maxR_2_obj}\\
    &\text{s.t.} & \cfrac{1}{\gamma_{k}} \leq \cfrac{1}{\gamma_{\rm min}}, \enspace \forall k=1,2, \ldots, K_2,\label{p1_throu:c1_2}\\
    && \eqref{p1_throu:c2}-\eqref{p1_throu:c7}. \label{p1_throu_2:c2} 
    \end{IEEEeqnarray*}
	
	Now, the objective as well as the constraints of problem (\ref{eq:maxR_2}) are posynomials.\footnote{The term `posynomial’ refers to a function consists of a sum of positive polynomials \cite{boyd2007tutorial}.} Therefore, optimal solutions can be found in polynomial-time using geometric programming (GP) approach \cite{boyd2007tutorial}.
	
	\subsection{Minimum Rate Maximization (Max-Min Rate)}
	By introducing a new slack variable $t$, (\ref{eq:max_min_R_general}) can be reformulated as
	\begin{IEEEeqnarray*}{lcl}\label{eq:max_min_R_2}
    &\underset{\mathrm{\Phi}_{k},N_{k},P_{k}^{\rm CS}}{\min} \, \, & t  \IEEEyesnumber \IEEEyessubnumber*\label{eq:max_min_R_2_obj}\\
    &\text{s.t.} & \cfrac{1}{\gamma_{k}} \leq \cfrac{1}{t}, \enspace \forall k=1,2, \ldots, K_2, \label{p2-2:c1} \\
		&& \eqref{p1_throu:c1_2}, \eqref{p1_throu:c2}-\eqref{p1_throu:c7}. \label{p2-2:c3} 
    \end{IEEEeqnarray*}
	The objective function (\ref{eq:max_min_R_2}) is a monomial function, whereas the constraints (\ref{p2-2:c1}-\ref{p2-2:c3}) are posynomial constraints. Therefore, it is an GP problem that can be solved optimally \cite{boyd2007tutorial}.
	
	\subsection{Reflecting Units Minimization}\label{Sec:minN_solut}
	Following the same procedure applied in the previous subsection, and by relaxing $N_{k}$ to be a continuous variable, the problem (\ref{eq:min_N_general}) can be re-written as
		\begin{IEEEeqnarray*}{lcl}\label{eq:min_N_21}
    &\underset{\mathrm{\Phi}_{k},N_{k},P_{k}^{\rm CS}}{\min} \, \, & \sum_{k=1}^{K_2}  N_{k}  \IEEEyesnumber \IEEEyessubnumber*\label{eq:min_N_2}\\
    &\text{s.t.} & \eqref{p1_throu:c1_2}, \eqref{p1_throu:c2}-\eqref{p1_throu:c7}. 	\label{p2_minN_2:c2}
    \end{IEEEeqnarray*}
	It is easy to notice that \eqref{eq:min_N_21} is also an GP problem and can be solved optimally \cite{boyd2007tutorial}, and the final value of $N_{k}$ is approximated as $\lceil N^*_{k} \rceil$.

	\begin{figure}[h]
		\centering
		\includegraphics[scale=0.9]{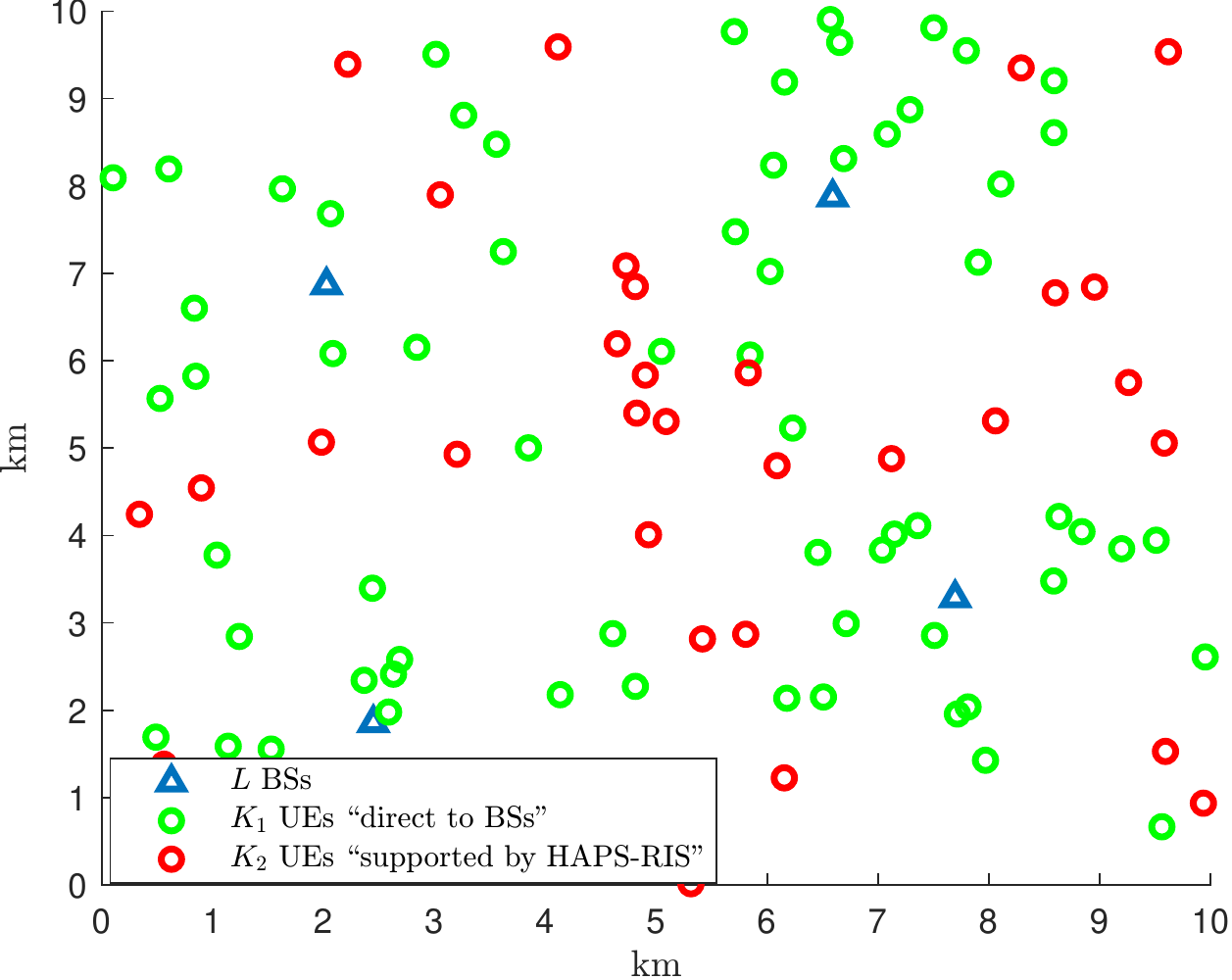}
		\caption{\textcolor{black}{Locations of the UEs $ \in \{\mathcal{K}_1, \mathcal{K}_2\}$ and BSs ($f_c=2$ GHz, $\sigma_{\rm BS-UE}=$ 8 dB).}}
		\label{fig:connection} 
	\end{figure}
	
	\section{Numerical Results and Discussion} \label{Sec:Results}
	
	In this section, we  discuss the performance of the proposed \textit{beyond-cell} communication approach by comparing  different power and RIS-unit allocation strategies obtained by solving the aforementioned problems (i.e., \eqref{eq:sum_rate_max}, \eqref{eq:max_min_R_general}, and \eqref{eq:min_N_general}) 
and a benchmark \textit{proportional} scheme. The benchmark scheme allocates the reflectors to each UE proportionally based on its channel gain, i.e., the UE with the worst channel gain will get the largest portion of the reflecting units.
	
	In the simulation setup, we consider an urban environment with an area of 10 km by 10 km consisting of \textcolor{black}{$L_{\rm max} = 4$} terrestrial BSs serving $K=100$ randomly and uniformly distributed UEs with a minimum separation distance of 100 m between the UEs. The BSs are typically placed where the UE density is expected to be higher. Therefore, the BS locations are optimized using 
   \textcolor{black}{Lloyd's 
	algorithm, which minimizes the distances between all the UEs and their associated BSs \cite{lu2016statistical}}.
	\textcolor{black}{Further, we adopt the 3GPP standards \cite{3gpp38901study} for terrestrial BS parameters with a BS height of $H_{\rm BS} = 25$ m and a  power and antenna gain of $P^{\rm BS} = 35$ dBm and  $G^{\rm BS} = 8$ dB. Also, we consider the communication at carrier frequency $f_c=2$ GHz with shadowing standard deviation $\sigma_{\rm BS-UE}=8$ dB, and all the UEs have the same height of 1.5 m. 
	Then, by following the urban channel model for the path loss and the LoS probabilities  detailed in \cite[Tables 7.4.1-1 -- 7.4.2-1]{3gpp38901study}, the channel gains between all UEs and all  BSs are obtained.}
	Unless stated otherwise, we consider $R_{\rm min} = 2$ Mb/s as the minimum rate for direct connection between a UE and a BS. 
\begin{figure}[h]
		\centering
		\includegraphics[scale=0.8]{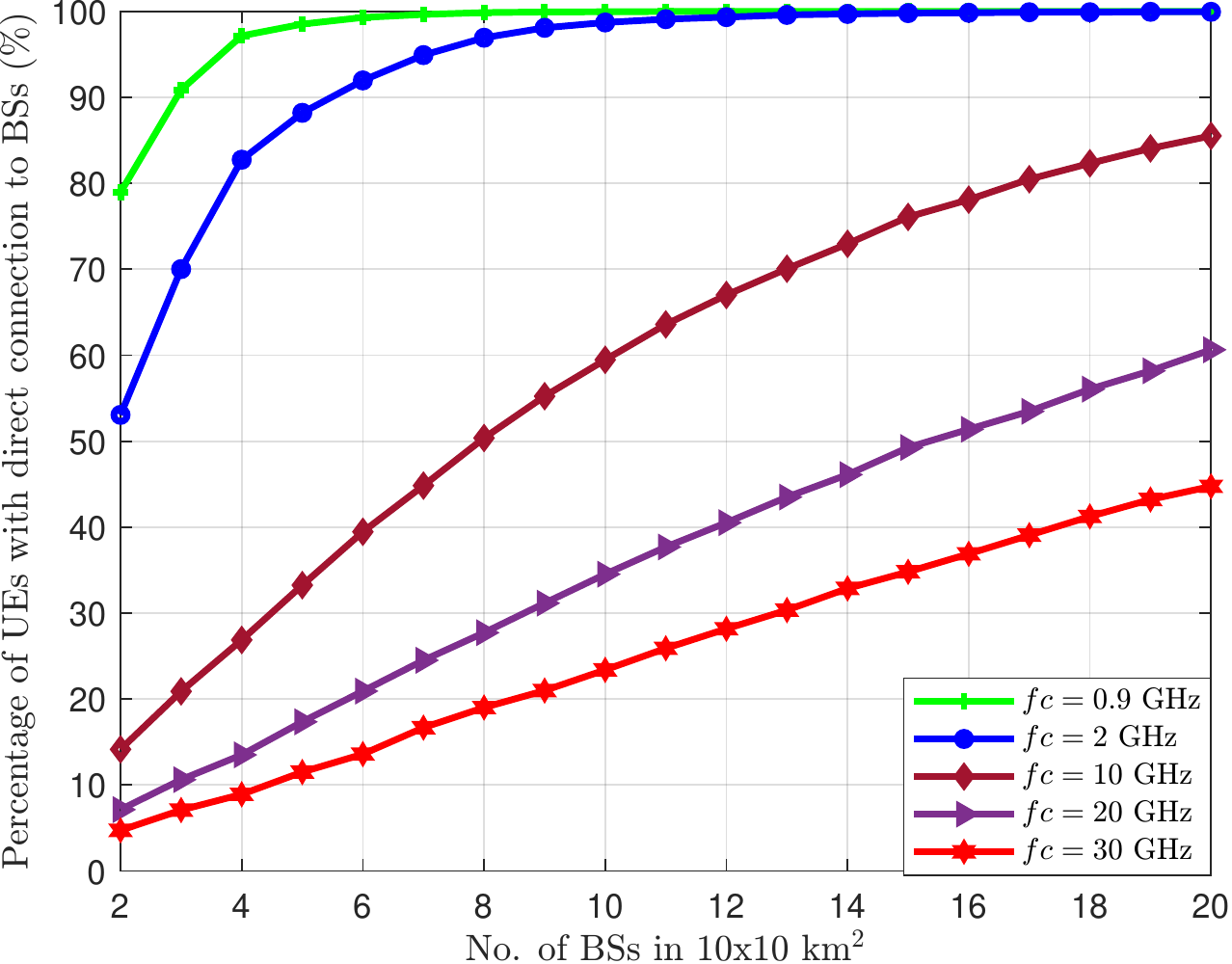}
		\caption{Relation between BS  densities and  percentage of UEs with direct connections for different frequencies.}
		\label{fig:connection_diff_freq} 
	\end{figure}
	
			\begin{figure*}[!ht]
		\begin{subfigure}{0.50\textwidth}
			\centering
			\includegraphics[width=\linewidth]{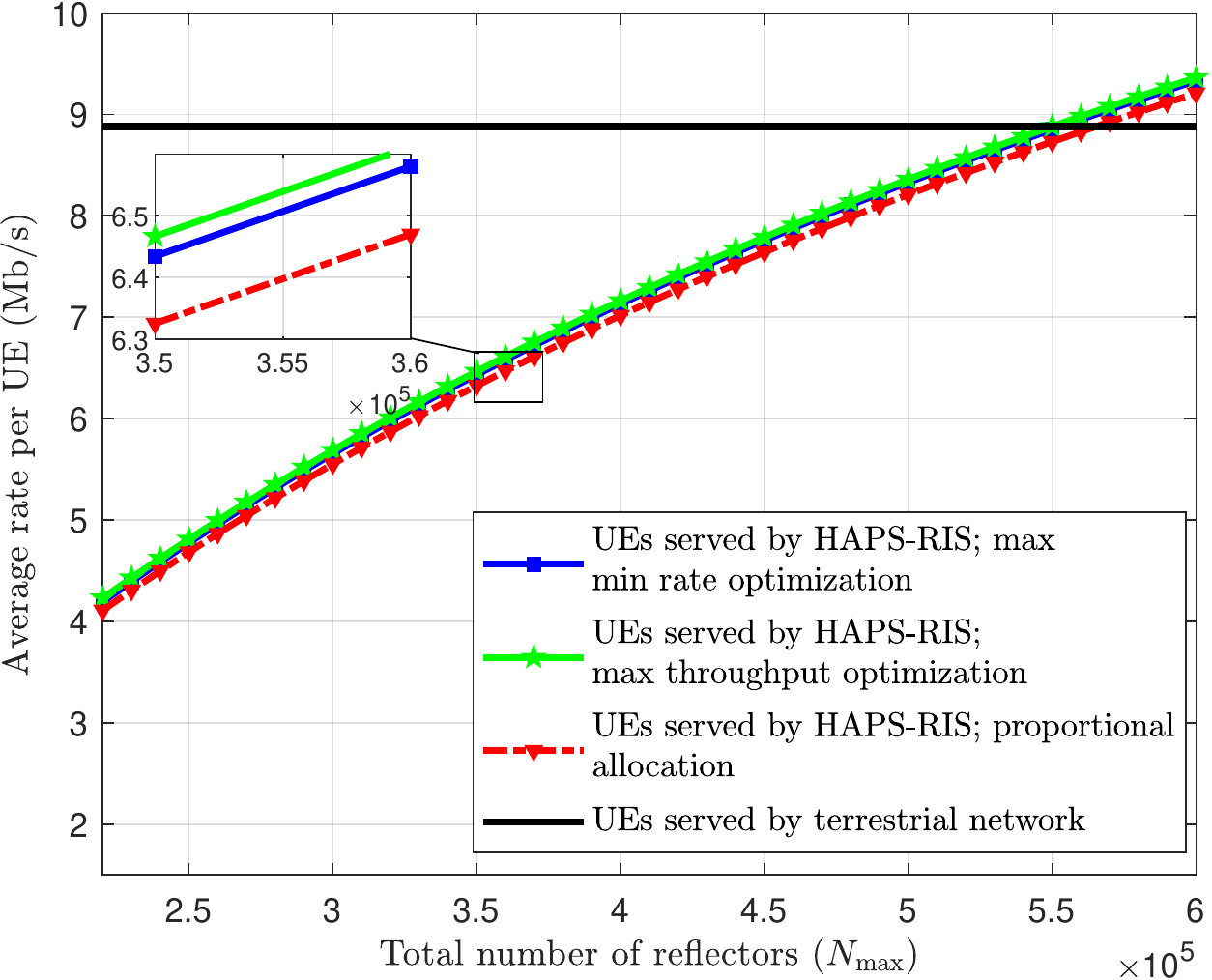}
			\caption{Average UE rate.}
			\label{fig:sfig1}
		\end{subfigure}
		\hspace{1em}
		\begin{subfigure}{0.5\textwidth}
			\centering
			\includegraphics[width=\linewidth]{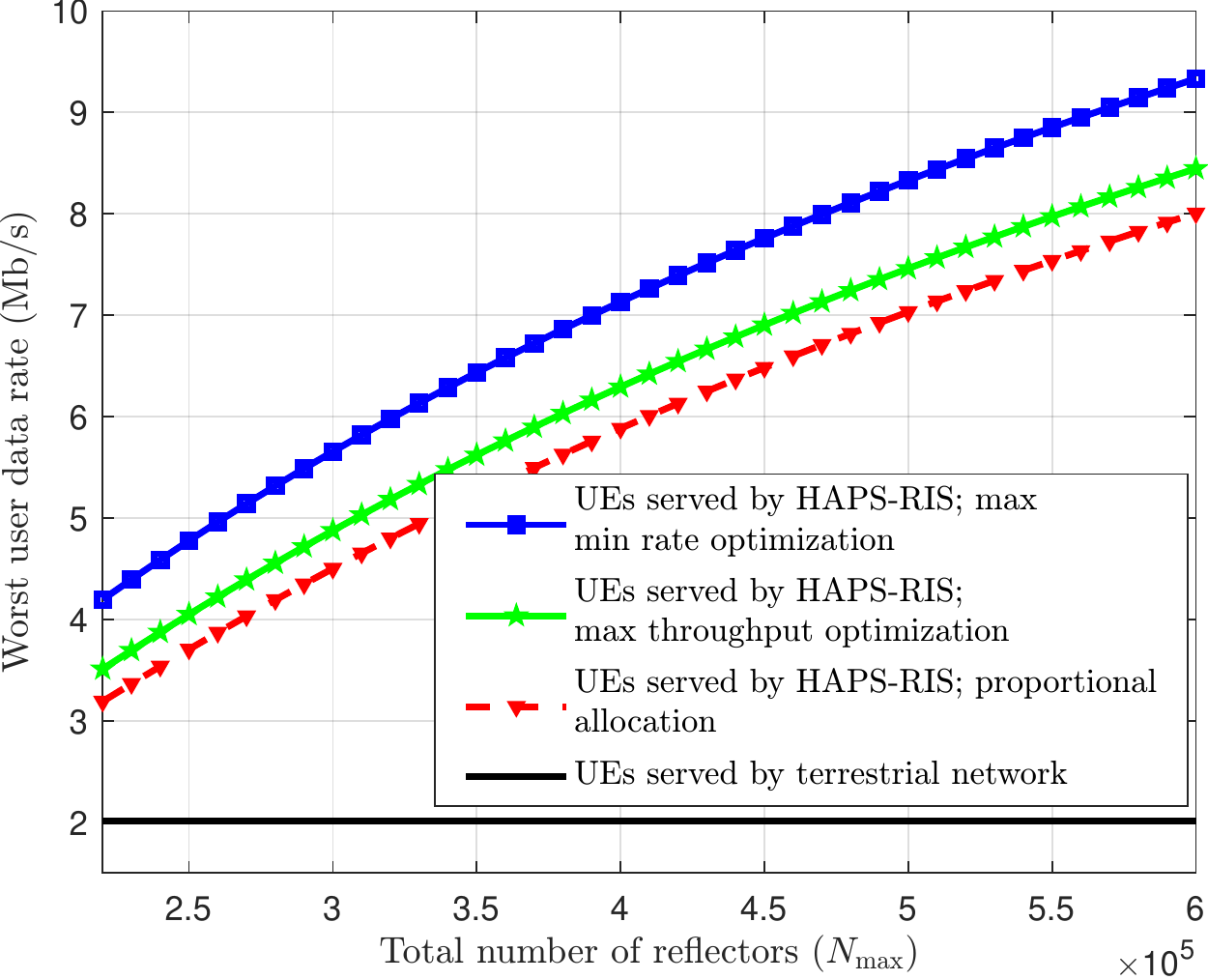}
			\caption{Worst UE rate.}
			\label{fig:sfig2}
		\end{subfigure}
		\caption{Comparison between terrestrial and HAPS-RIS communications with different allocation strategies.}
		\label{fig:opt_comp}
	\end{figure*}

		\begin{figure}[h]
		\centering
		\includegraphics[scale=0.90]{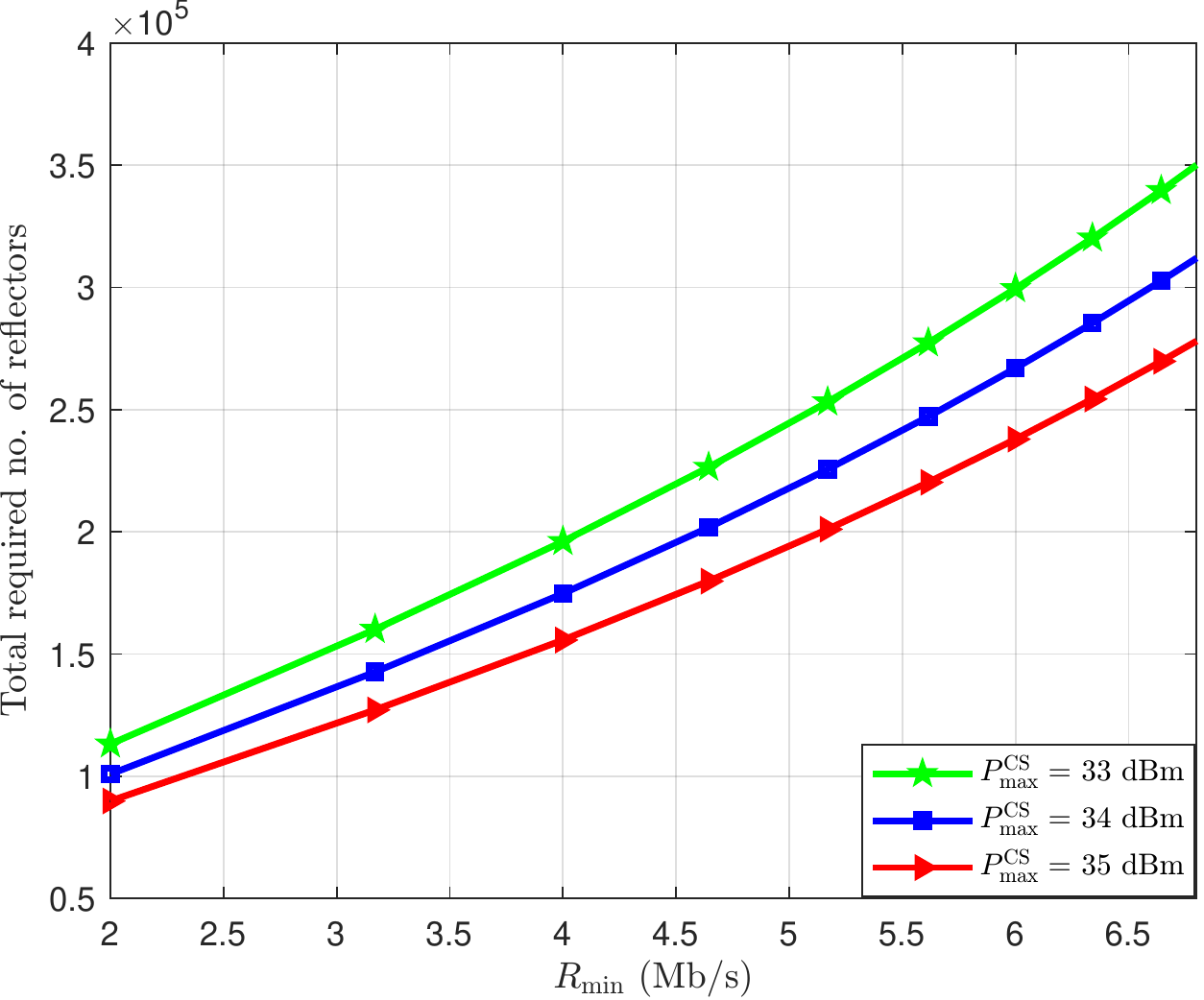}
		\caption{Relation between $R_{\rm min}$ for 
		 HAPS-RIS served UEs and the total required number of reflecting units.}
		\label{fig:R_min_vs_N} 
	\end{figure}
	
	Accordingly, a UE will be associated with a BS that provides the highest data rate. Fig.~\ref{fig:connection} illustrates the optimized locations of BSs among randomly and uniformly distributed UEs. The UEs marked with red circles do not satisfy the minimum rate requirement for any BSs. Therefore, they will be served by the CS through HAPS-RIS. \textcolor{black}{Following the standardized 3GPP channel model 
		established for a HAPS and terrestrial nodes in urban environments
		 \cite[Sec. 6]{3gpp2017Technical}, and the scattering reflecting paradigm of the RIS as detailed in \cite{alfattani2021link}, the effective channel gains from the CS to all  UEs in set  $\mathcal{K}_2$ through the HAPS-RIS are obtained (i.e., $\textsf{PL}
^{\rm CS- HAPS}$ and $\textsf{PL}
^{\text{HAPS-}k}$).} For this model, we consider dry air atmospheric attenuation.  The atmosphere parameters are selected on the basis of the mean annual global reference atmosphere \cite{itu1999p}. Further, we assume the UEs have 0 dB gain, $N_0 = -174$ dBm/Hz, $B_{\rm BS}=50$ MHz  and $B_{\rm UE}=2$ MHz. We further set  $P_{\rm max}^{\rm CS} =$  33 dBm, and $G_{\rm t}^{\rm CS}=$ 43.2 dB \cite{3gpp2017Technical} in all of the simulations, unless otherwise stated. 

	\subsection{Relationship between BS Density and Direction Connection with UEs}
	It should be noted that the percentage of UEs supported by the CS via the HAPS-RIS depends on the number of UEs that fail to connect
	 with any terrestrial BS directly. 
	The chances of UEs being supported by the BSs directly 
	depends on the BS and UE densities and the carrier frequency.
	 For a fixed density of BSs, as the density of UEs increases, the percentage of UEs with direct connections drops because the BSs cannot serve more users beyond their maximum loading capacities.
	Fig.~\ref{fig:connection_diff_freq} 
	shows how the percentage
	 of UEs with direct connections increases as the density of BSs increases.
	However, as the carrier frequency increases to provide high data rate communications, the percentage of UEs with direct connections significantly drops, even with the large number of BSs.
	For instance, four BSs communicating at $f_c \leq 2$ GHz are sufficient to support more than 80\% of UEs directly connected to BSs, whereas more than 20 BSs are required to support 80\% of UEs communicating at $f_c \geq 10$ GHz. Therefore, the HAPS-RIS may 
	offer a more cost-effective solution in such situations than
	 just increasing the density of the BSs.

	\subsection{Comparison Between Allocation Schemes} 
	\subsubsection{Sum rate and worst UE rate maximization based allocation}
	In this simulation,  we set $N_{k, \rm min}=1000$ , $N_{k, \rm max}=10,000$ reflectors and $P_{k, \rm min}^{\rm CS}=15$ dBm and $P_{k, \rm max}^{\rm CS}=20$ dBm. 
	
	\textcolor{black}{Fig. \ref{fig:opt_comp} compares the achievable average and worst rate performances for the UEs belonging to set $\mathcal{K}_1$ and $\mathcal{K}_2$. For set $\mathcal{K}_2$, the performance plots are obtained using the optimized power and reflecting units allocation schemes and are compared with the \textit{proportional} allocation strategy for benchmark purposes. }
	
	\textcolor{black}{For the selected range of $N_{\rm max}$, i.e., 200,000 -- 600,000 reflecting units\footnote{\textcolor{black}{\textcolor{black}{This is equivalent to an RIS area between 180 - 540 $\rm m^2$. } This can represent a partial area of a HAPS, as the length of an airship is
between 100 and 200 m,  whereas an aerodynamic HAPS has wingspans between
35 and 80 m. However, the size of each reflector unit is about $(0.2\lambda)^2$ \cite{kurt2021vision}.}}, it can be observed that for most of the values of $N_{\rm max}$ the average rate of the terrestrial UEs is  higher than that of the UEs supported by the HAPS-RIS (Fig. \ref{fig:sfig1}). This is because the average performance is dominated by the excellent channel conditions between some UEs belonging to set $\mathcal{K}_1$ and the BSs. However, 
\textcolor{black}{for HAPS-RIS to outperform RIS-assisted terrestrial networks in terms of the average rate, $N_{\rm max}$ should be more than 550,000 reflecting units. }\textcolor{black}{Because the rate performance of the worst UE is one of the concerns of network operators, we see in Fig. \ref{fig:sfig2} that the rate performance of the worst UE assisted by HAPS-RIS is significantly higher than that are supported by the terrestrial networks.}}
	
	\textcolor{black}{For the UEs belonging to set $\mathcal{K}_2$, we observe that the \textit{max throughput} allocation scheme achieved the best performance in terms of  average UEs rate. }
	However, in terms of the worst UE performance, the \textit {max-min R} allocation scheme significantly improves the rate of the UE with the weakest channel gain, and it substantially outperforms the \textit{max throughput} and the \textit{proportional allocation} schemes. It should be noted that the improvement in the worst UE rate leads to the degradation of the sum and average UE rates. Since the \textit{max min R} scheme distributes the system resources fairly and maximizes the fairness among all the UEs, it results in a performance loss for the whole system. 
	\textcolor{black}{It is interesting  to note from Fig. \ref{fig:opt_comp} that the }
	performance enhancement for the worst UE rate by \textit{max min R}  scheme is about 15\%, while the degradation in terms of the average rate or  throughput is 1\% less than  the optimized \textit{max throughput} allocation scheme.

	\subsubsection{Reflectors Minimization Based Allocation}
	Fig. \ref{fig:R_min_vs_N} shows the variation of the minimum number of reflectors required with the different values of the minimum rate requirements of the UE. The number of reflectors and power $P_{k}^{\rm CS}$ corresponding to all $\mathcal{K}_2$ UEs that satisfy the minimum rate requirements are obtained by solving the problem \eqref{eq:min_N_2}. As we can see, an almost linear relationship exists between the rate requirement and the  minimum required number of reflectors. Moreover, by doubling the rate required for the UEs, the RIS unit requirement is increased by 100\%.   Fig. \ref{fig:R_min_vs_N} also shows the relationship between the different values of the maximum transmit power available at the CS ($P_{\rm max}^{\rm CS}$) and the optimized number of reflectors. It can be observed that by increasing  $P_{\rm max}^{\rm CS}$ by 1 dB, the minimum required  number of reflectors is reduced by about 11\%.

	\addtolength{\topmargin}{0.01in}
	\section{Conclusion}\label{Sec:conclstion}
	In this paper, we introduced the novel concept of \textit{beyond-cell} communications using HAPS-RIS technology to complement  terrestrial networks by supporting  unserved UEs. We formulated three resource allocation optimization problems to design the CS power and RIS unit allocation strategies. 
	The optimization objectives included throughput maximization, max min rate, and minimal usage of RIS reflectors. 
	The results showed the capability of  \textit{beyond-cell} communications approach to 
	support a larger number of users with a minimal number of terrestrial BSs. Furthermore, the results showed the superiority of the proposed solutions over the benchmark approach and demonstrated the trade-off between the total and average rate performance and the fairness among UEs. 
	
	\bibliographystyle{IEEEtran}
	\bibliography{IEEEabrv,Final_Globecome_workshop_camera_ready}
\end{document}